\documentclass{aa}
 \usepackage{psfig,times}

\newcommand{\be}{\begin{equation}}
\newcommand{\ee}{\end{equation}}
\newcommand{\bea}{\begin{eqnarray}}
\newcommand{\eea}{\end{eqnarray}}

\def\apj{ApJ}
\def\apjs{ApJS}
\def\apjl{ApJL}
\def\aea{A\&A}
\def\aear{A\&A Rev.}

\def\aas{A\&AS}

\def\la{\mathrel{\hbox{\rlap{\hbox{\lower4pt\hbox{$\sim$}}}\hbox{$<$}}}}
\def\ga{\mathrel{\hbox{\rlap{\hbox{\lower4pt\hbox{$\sim$}}}\hbox{$>$}}}}

\newcommand{\gi}{$\chi^2$}

\def\la{\mathrel{\hbox{\rlap{\hbox{\lower4pt\hbox{$\sim$}}}\hbox{$<$}}}}
\def\ga{\mathrel{\hbox{\rlap{\hbox{\lower4pt\hbox{$\sim$}}}\hbox{$>$}}}}

\def\aea{A\&A}
\def\apj{ApJ}
\def\apjs{ApJS}

\begin{document}

\title{The X-ray spectra of the flaring and quiescent states of AT~Microscopii observed by XMM-Newton \thanks{Based 
on observations obtained with {\sl XMM-Newton}, an ESA scienc mission with instruments and contributions directly funded by
ESA Member States and the USA (NASA)} 
}

\author{A.J.J.\,Raassen\inst{1,2} \and R.\,Mewe\inst{1} \and M.\,Audard\inst{3} \and M.\,G\"udel\inst{4}}

\institute{
SRON National Institute for Space Research, Sorbonnelaan 2, 3584 CA Utrecht, The Netherlands 
\and 
  Astronomical Institute Anton Pannekoek, Kruislaan 403, 1098 SJ Amsterdam, The Netherlands
\and
 Columbia Astrophysics Laboratory, Mail code 5247, 550 West 120th Street,
 New York, NY 10027, USA
\and
  Paul Scherrer Institut, W\"urenlingen \& Villigen, 5232 Villigen PSI, Switzerland }

\offprints {A.J.J.Raassen,\\
 \email a.j.j.raassen@sron.nl}
\authorrunning{A.J.J.Raassen, R.Mewe, M.Audard et~al.}
\titlerunning{XMM-Newton X-ray spectrum of AT Mic}

\date{Received date \today; accepted date $\ldots$}


\abstract{
The X-ray spectrum of the late-type M-dwarf binary AT Mic (dM4.5e+dM4.5e) is observed in the wavelength range 1 - 40 \AA\ by
means of {\sc rgs} and {\sc epic-mos} on board XMM-Newton. During the exposure a flare occured. 
We have performed a 3-temperature fit and a DEM-modeling to the flaring and quiescent part of the spectrum. 
We report the coronal temperature distribution, emission measures, and abundances of the flaring and quiescent state 
of this bright X-ray source. The temperature range stretches from about 1 to 60 MK. The total volume emission measure in 
this temperature interval is $\sim 12.2 \times 10^{51}$ cm$^{-3}$ for the quiescent state and $\sim 19.5 \times 10^{51}$ cm$^{-3}$ 
for the flare state. This difference is due to the contribution of the hot temperature component.
The high-resolution spectrum of AT Mic, obtained by {\sc rgs}, is dominated by the H- and He-like transitions 
of C, N, O, and Ne and by  Fe~{\sc xvii} lines, produced by the plasma with temperatures from 1 to 10~MK. 
The {\sc epic-mos} spectrum below 10~\AA\ shows H- and He-like 
Ne, Si and the iron K-shell transitions. They are produced by the hot component (30~MK). The iron K-shell is more
prominent in the flare state.
The abundance pattern in the quiescent state of AT Mic shows the depletion of low-FIP elements relative to 
high-FIP elements, indicating the presence of an I(nverse)FIP effect in this active star. In the flare state, however,
some flattening of this IFIP effect is present.
}

\maketitle

\section{Introduction}
  
Hot outer atmospheres (coronae) are very common for relatively cool stars in the
spectral classes F-M. 
Many of these coronae are characterized by 
temperatures up to about 20 MK and densities $n_\mathrm{e}\ga10^{10}$~cm$^{-3}$. It was already known 
that the coronae of many stars are different from the solar corona (e.g., reviews by Pallavicini 1989 and Mewe 1991). The
heating mechanism of these coronae, however, is still not well understood. A variety of possible explanations is present, such as,
convection zones with MHD waves transfering energy from the photosphere into the corona, 
large flares heating plasma by reconnection, or a large number of continuously heating micro- and nanoflares.
 High-resolution spectroscopy of stellar systems available from Chandra (Brinkman et al. 2000,  Canizares et al. 2000) and XMM-Newton 
(Brinkman et al. 2001) offers  the possibility to study the coronal spectra in great detail and to determine 
various coronal quantities, 
such as densities, temperatures, abundances, emission measures, and line ratios, 
during the flaring states as well as during the quiescent state.

Late-type stars (M and K dwarfs) show more magnetic activity and higher
coronal temperatures than the Sun,
 characterized by flaring (Pettersen 1989) in the optical, UV, X-ray and radio wavelengths (Pallavicini et al. 1990). 
To understand the mechanisms that underly the high temperatures and the variety of densities, emission measures,
and abundance values observed in coronal systems, the X-ray spectra of stars of type F-M are investigated, 
especially focused on stars with flare activity. In very active stars, however, the ``quiescent state" might 
be produced by a superposition of flare decays and microflares rather than by a real quiescent coronal plasma. 
Here we present the investigation of the active M-type binary star AT Mic (dM4.5+dM4.5)
at a distance of 10.22 pc (Perryman et al 1997).
Both components are known to flare frequently (Joy \& Wilson, 1949). Large optical flares were
observed by Garc\'\i a-Alvarez et al. (2002), while X-ray flares were studied e.g., by HEAO-1 (Kahn et al. 1979) and by EXOSAT (Pallavicini et al. 1990). 
The quiescent corona was observed by EUVE (Monsignori Fossi et al. 1995).

\section{Observations}

The X-ray spectrum of AT Mic was observed by {\sc rgs} and {\sc epic-mos} on board XMM-Newton on 14 October 2000 
during revolution 156.
The observation log of the data is presented in Table~1.

\begin{table*}[ht]
\caption{Observation log of the data of AT Mic.}
\begin{center}
\begin{tabular}{|l@{\ }|l@{\ }|l@{\ }|l@{\ }|l@{\ }|r@{\ }|}
\hline
Instrument           &  Filter&Mode&Date-obs-start&Date-obs-end&Performed duration(sec)\\
\hline
MOS1                 &  Medium&Small Window&2000-10-16T00:27:48&2000-10-16T07:34:02&25394\\
MOS2                 &  Medium&Timing      &2000-10-16T00:38:15&2000-10-16T07:34:29&24794\\
pn                   &  Medium&Small Window&2000-10-16T00:41:50&2000-10-16T07:41:13&25103\\
RGS1                 &  None  &Spec+Q      &2000-10-16T00:19;12&2000-10-16T08:11:48&28254\\
RGS2                 &  None  &Spec+Q      &2000-10-16T00:19:12&2000-10-16T08:11:48&28254\\
\hline
\end{tabular}
\end{center}
\end{table*}

The data were processed by means of the XMM-Newton SAS version 5.3.3 (June 2002). 
The exposure time was 28.3~ks for {\sc rgs} and 25.4~ks for {\sc epic-mos1}, of which
19.0~ks were free of solar flare proton ``pollution" applying the criterion that the latter's
count rate should be $\la$ 0.35~c/s. 
For {\sc rgs} the
first-order net spectra were extracted by including 95\% of the cross dispersion PSF (xpsfincl = 95 in rgsproc)
and the background spectra were extracted by excluding 98\% (xpsfexcl = 98 in rgsproc).

The {\sc rgs} spectral resolution 
is $\Delta\lambda\sim$0.07 \AA\ (FWHM), with a maximum effective area of about 140 cm$^2$ around 15~\AA.  
The wavelength uncertainty is 7-8 m\AA. 
The total {\sc rgs} spectrum runs from 5 to 37~\AA. Only the range from 8 to 37~\AA\
was used. For {\sc epic-mos}1, which observed in the small-window mode, we extracted the spectrum by means
of a circle centered around the source with a radius of 40~arcsec. The background was obtained by applying
the same circle in a source free area at another CCD. The same SAS version was used to construct 
response matrix and ancillary response files.  The range from 1 to 14~\AA\ was used.
{\sc epic-mos}2 was operating in the timing mode. 
For more instrumental details on {\sc rgs} and {\sc epic-mos} see 
den Herder et al. (2001) and Turner et al. (2001), respectively. 
The pn~data were used to produce a lightcurve only.

\section{Analysis}
\subsection{Quiescence versus flaring state}
\begin{figure}[h]
\hbox{
\psfig{figure=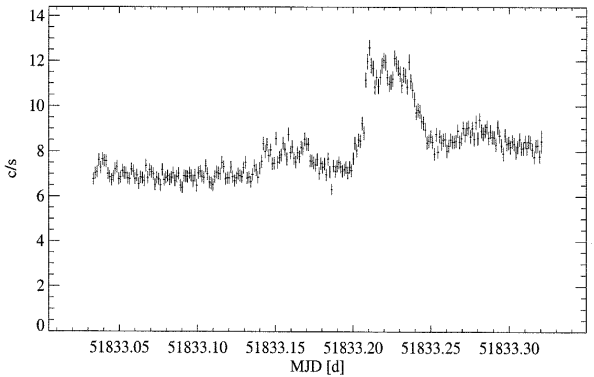,angle=-90,height=5truecm,width=8.0truecm,clip=}}
\caption{Lightcurve of AT Mic, observed with pn.}
\end{figure}

Based on the lightcurves obtained with RGS, EPIC-MOS and pn (see Fig.~1) we have concluded that AT~Mic
was flaring several times during our observations. A period of about 2.9~ks related to the most prominent 
flare was extracted from the GTI table as flare state, leaving a pseudo-quiescent part of 14.1~ks. Three 
other weak flares can be seen in Fig.~1. However, they do not dominate and
have therefore not been flagged as flares in the further spectral analysis.
From Fig.~1 we notice that the top of the flare is not constant, but shows small variations, which seem to be periodic.
However, these time intervals are too short to allow for individual spectral analysis.

\begin{figure*}
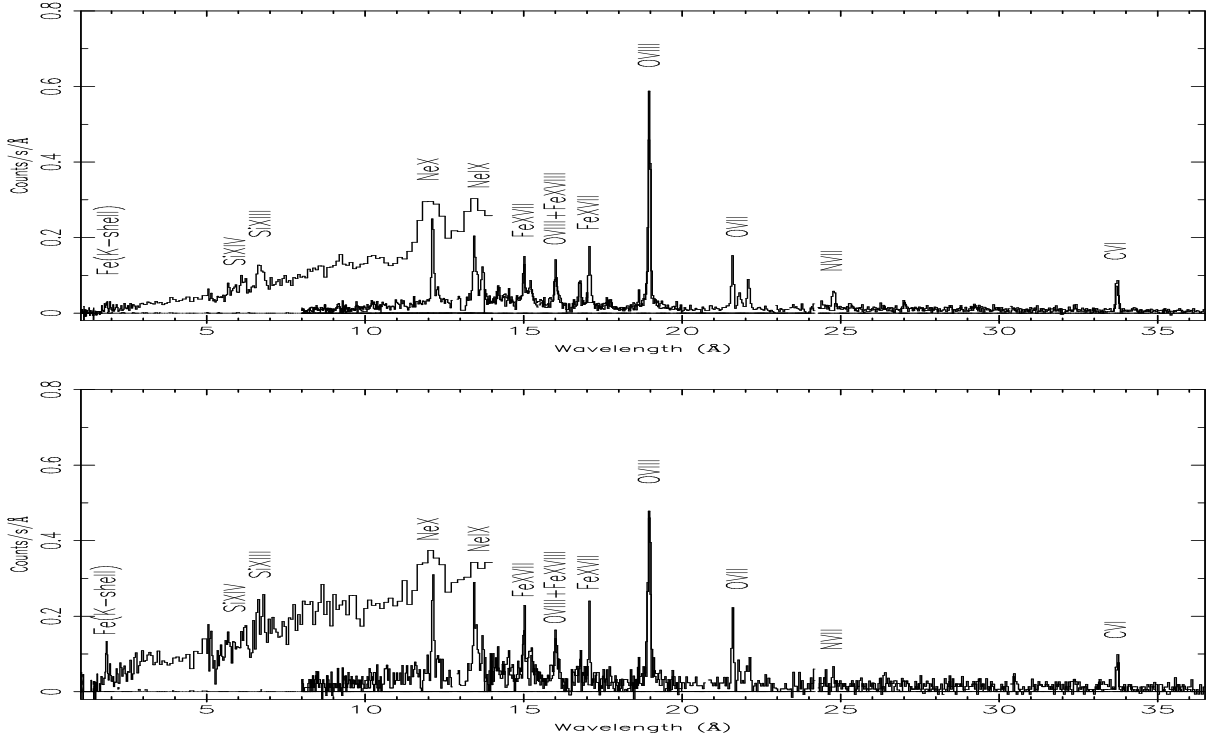

\hbox
{\psfig{figure=h4290F2a.ps,angle=-90,height=5truecm,width=16.0truecm,clip=}}
{\psfig{figure=h4290F2b.ps,angle=-90,height=5truecm,width=16.0truecm,clip=}}
\caption[]{ The spectra of AT Mic observed by {\sc rgs}1, {\sc rgs}2 and {\sc epic-mos} in the wavelength region from 1 to 37 \AA. 
The upper panel shows the quiescent state, while the lower panel shows the flaring state. }  
\end{figure*} 

The observed spectra are shown in Fig.~2 for the quiescent (top) and flaring (bottom) state. 
As can be seen from Fig.~2, {\sc epic-mos} has a low resolution, but a high sensitivity compared to {\sc rgs}. 
Due to the short exposure time of the flaring state the signal/noise ratio is lower for that spectrum. 
For both states shown in Fig.~2 it is clear that the X-ray spectrum obtained by {\sc rgs} in the wavelength range from 5 to 37~\AA\ is 
dominated by H- and He-like transitions of C, N, O, and Ne and by Fe~{\sc xvii} lines, while the low-resolution
{\sc epic-mos} spectrum shows the Fe K-shell transitions around 1.9~\AA\ and the H- and He-like 
transitions of Ne and Si. The Fe K-shell feature (see also Table~4) as well as the
continuum appear more prominent in the spectrum of the flaring state (bottom panel) than in that of the quiescent state.

\subsubsection{Multi-temperature fitting}
The data of the quiescent state and the flaring state have been fitted with a 3-$T$ Collisional Ionization Equilibrium model using 
the {\sc spex} code (Kaastra et al., 1996a) in combination with an updated version of {\sc mekal} 
(Mewe et al. 1995, Phillips et al. 1999). 
The {\sc mekal} database is publicly available\footnote
{\textrm{http://www.sron.nl/divisions/hea/spex/version1.10/line/line\_new.ps.gz} }
 as an extended list of fluxes of more than 5400 spectral lines. 
Fig.~3 shows the comparison of the quiescent spectrum with the best-fit model. The fits have been performed by
attaching the errors to the data (``fit weight data"), given in the columns ``data", and by attaching the errors
to the model  (``fit weight model"), given in the columns ``model". In the abstract the values from the ``data fit"
are given.

\begin{figure*}
\hbox{
\psfig{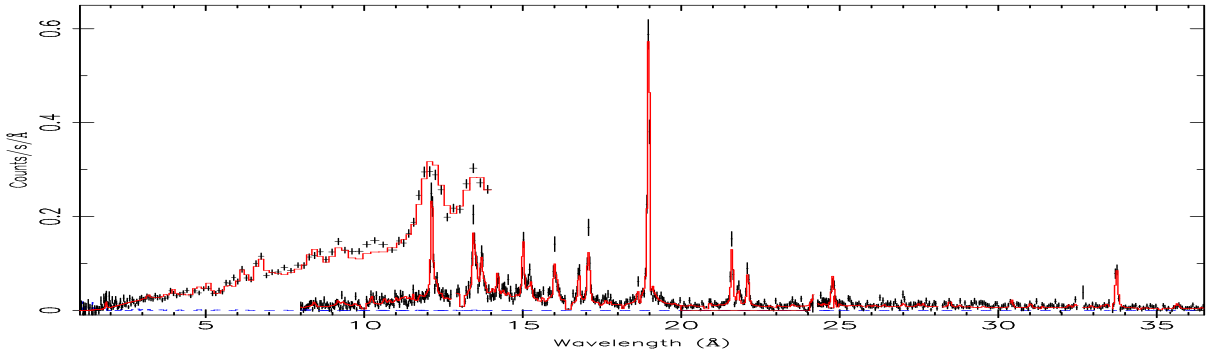}}
\caption{Comparison of the AT~Mic quiescent spectrum (see Fig.~2a) with the best-fit model (unbroken line).}  
\end{figure*}

Table~2 shows the 3~temperatures, the emission measures in 10$^{51}$~cm$^{-3}$, and abundances together 
with their standard deviations within parentheses.
The abundances are normalized to their 
corresponding solar photospheric values (Anders \& Grevesse 1989), except for iron and oxygen.
For Fe we have changed the value of log A$_{\rm Fe}$ from 7.67 (Anders \& Grevesse 1989) 
to 7.51 (Grevesse \& Sauval 1998 and 1999).  For oxygen we used the new value given by Allende Prieto \& Lambert (2001)
(log A$_{\rm O}$ = 8.69 instead of 8.93).
Here log A$_{\rm Fe}$ and log A$_{\rm O}$ is the logarithm of the Fe- and O-abundance relative to log A$_{\rm H}$=12.0.

The temperatures in Table~2 cover a range from 3 to 35 MK. It turns out that the line spectrum is mainly 
produced by the EM related to $T_1$~=~3~MK and to $T_2$~=~8~MK, while the hot component $T_3 >$20~MK 
contributes to the continuum below 10~\AA\ and the Fe K-shell lines in
the {\sc epic-mos} observation. This is especially true for the flaring state.
A fourth cool component (1~MK) is possible but not well constrained. This cool component is only related to the {\sc C vi} line
(influencing the C abundance) and very weakly coupled to the continuum. 
The total emission measure of the three components is 12.2(.5)$\times10^{51}$~cm$^{-3}$ for the quiescent state
and 19.5(.8)$\times10^{51}$~cm$^{-3}$ for the flare state.
Based on EUVE observations, Monsignori Fossi et al. (1995) have determined an emission measure maximum of about
3$\times10^{51}$~cm$^{-3}$ around  10~MK (their Figure~5), detecting only the emission measure related to
our second temperature component at $T_2$(7.9~MK). 

When fitting the O~VII lines a slight increase of the  electron density $n_e$ is noticed, 
from the quiescent state to the flaring state. However, this increase is not statistically significant.

In the energy interval from 0.3 to 10~keV the X-ray luminoscity $L_x$ is 2.13(.08)$\times10^{29}$erg/s for 
the quiescent state and
3.63(.22)$\times10^{29}$erg/s for the flaring state. These values are averages of the ``data" and ``model" fit.
Pallavicini et al. (1990) observed AT Mic with EXOSAT (ME 1-10~keV detector)
during quiescence and during a flare. The quiescent X-ray luminosity was measured to be 1.8~10$^{29}$~erg/s
(Table~5 of Pallavicini et al. 1990) in agreement with our value.
However, they observed a larger flare, but with the same temperature as ours.

The abundances have been determined by fitting to the spectrum as a whole. As a check, whether a
large scatter exists between different lines of the same element, abundances were also fitted to isolated 
individual lines, keeping temperatures, emission measures and other abundances fixed 
at the values from the global fit (Table~2). In this way blends have been taken into account. 
Due to the weak continuum the absolute abundances cannot be coupled accurately to the continuum and 
therefore they are strongly anticorrelated with the emission measure. 
This means that if one quantity increases the other decreases and vice versa. This is clear from comparing the
$EM_{\rm total}$ from the ``data fit" with that from the ``model fit" and by comparing the absolute Oxygen abundance (O/H)
in these columns. The product $EM_{\rm total}$$\times$ O/H, however, is more stable.
For this reason the abundance {\it ratios}, normalized to oxygen, are more robust. 
Oxygen was chosen for its low relative 
standard deviation, thanks to the strong O~{\sc viii} and O~{\sc vii} lines. 
For the (less active) solar corona Feldman et al. (1992) discussed the presence of a First Ionization Potential (FIP) effect.
This effect implies that elements with a low
value of the first ionization potential (e.g., $\la$ 10~eV) 
show enhanced coronal abundances relative to photospheric values.
However, the abundance ratios of the low-FIP elements Mg, Si, and Fe
are relatively low in AT~Mic, indicating an {\sl inverse} FIP-effect (IFIP). 
Similar IFIP effects were observed by Brinkman et al. (2001), G\"udel et~al. (2001), Audard et~al. (2002) and Audard et~al. (2003) for other active late-type stars. 
This IFIP effect is illustrated in Fig.~4 (top), which shows the trend of an {\sl increasing} abundance ratio with 
increasing FIP for the quiescent state. In the bottom panel the same quantities are shown for the flaring state. 
However, due to the short time interval
all quantities determined in the flare spectrum have large statistical uncertainties. 
The poor signal-to-noise ratio of the flaring state data is also reflected in the ``too good" \gi/dof 
of the multitemperature fit. This suggests an overestimate for the errors in this low count rate spectrum.
 For that reason the fits have also been performed, attaching the errors to the model, resulting in reasonable
\gi/dof-values.

Comparing the quiescent state with the flaring state some flattening of this IFIP effect for the low FIP elements 
(Fe and Mg) is noticed in the flaring state.
 This is shown in Table~3 and Fig.~4c, in which the ratios between the
A/O~values of the flaring and quiescent state are given. The values used in Table~3 and Fig.~4 are averages between
``data" and ``model" values. Si seems not to be affected by the difference between the flaring state and the quiescent state. 
In the solar corona the enhancement of the low-FIP elements starts from 10~eV down. 
Results in Table~3 and Fig.~4c may indicate a value lower than 10~eV below which the elemental abundances are influenced. 
Such a shift might be related to the lower surface temperature of AT~Mic.

\begin{table}[h]
\caption{Multi-temperature fitting to the quiescent and flaring state of AT Mic; 
1$\sigma$ uncertainties are given in the last digits in parentheses}
\begin{center}
\begin{tabular}[h]{|l@{\ }|l@{\ }l@{\ }|l@{\ }l|}
\hline 
\multicolumn{1}{|l|}{{\sc Parameter}}&\multicolumn{2}{c|}{{\sc quiescent}$^a$}&\multicolumn{2}{|c|}{\sc flaring}\\ 
\hline
                   &  data$^b$&  model$^b$&  data&  model\\
\hline
log $N_H$ [cm$^{-2}$] &18.3$^c$&18.3$^c$&18.3$^c$&18.3$^c$\\
$T_1$ [MK]            &3.16(.10)&2.94(.10)&2.91(.17)&2.71(.20)\\
$T_2$ [MK]            &7.77(.13)&7.54(.14)&8.57(.34)&7.80(.27)\\
$T_3$ [MK]            &23.0(1.0)&27.9(2.3)&33.9(3.2)&32.8(5.1)\\
$EM_1$ [10$^{51}$cm$^{-3}$]   &2.34(.16)&3.86(.41)&2.71(.31)&4.95(1.3)\\
$EM_2$ [10$^{51}$cm$^{-3}$]   &4.85(.26)&8.58(.93)&4.27(.42)&10.9(4.0)\\
$EM_3$ [10$^{51}$cm$^{-3}$]   &4.99(.20)&5.51(.44)&12.47(.61)&16.5(1.9)\\
$EM_{\rm total}$[10$^{51}$cm$^{-3}$]&12.2(.5)&17.8(1.1)&19.5(.8)&32.4(4.6)\\
$n_e$(O) [10$^{10}$cm$^{-3}$]$^d$&1.9(1.5)&1.4(1.4)&4(-3,+5)&3(-3,+4)\\
$L_x$ [10$^{29}$erg/s]$^e$&2.05(.08)&2.26(.14)&3.54(.15)&3.98(.57)\\
Mg/O \,7.65eV&0.51(.06)&0.44(.09)&1.00(.22)&0.68(.28)\\
Fe/O \,\, 7.87eV&0.34(.02)&0.31(.05)&0.56(.04)&0.45(.16)\\
Si/O \,\,\, 8.15eV&0.60(.06)&0.69(.11)&0.72(.22)&0.68(28)\\
C/O \,\, 11.26eV&1.00(.10)&1.01(.19)&0.83(.21)&0.84(.30)\\
O/O \,\, 13.62eV&1.00(.05)&1.00(.14)&1.00(.10)&1.00(.25)\\
N/O \,\, 14.53eV&0.86(.09)&0.73(.16)&0.58(.27)&0.51(.31)\\
Ne/O \, 21.56eV&1.52(.08)&1.48(.20)&1.90(.21)&1.50(.49)\\
O/H\,\,\,\, 13.62eV&1.57(.05)&0.96(.09)&1.43(.10)&0.71(.13)\\
$EM_{\rm total}$$\times$ O/H&19.2(1.0)&17.1(2.0)&27.9(2.3)&23.0(5.3)\\
\gi/dof &1.1&1.4&0.60&1.10\\
\hline
\end{tabular}
\end{center}
\begin{flushleft}
{
\begin{description}
\item $^a$ based on good time interval (with low solar proton flux) and excluding the stellar flare (see Fig.~5)
\item $^b$ data stands for ``fit weight data" and model stands for ``fit weight model"
\item $^c$ fixed at literature value from Monsignori Fossi et al. (1995)
\item $^d$ based on O~VII lines only
\item $^e$ in the 0.3--10 KeV energy interval 
\end{description}
}
\end{flushleft}  
\end{table}


\begin{table}[h]
\caption{Ratio between the scaled abundance values of the flaring and quiescent state
from Table~2}

\begin{center}
\begin{tabular}{|l|l|}
\hline
Parameter            &  flaring/quiescent\\
\hline
Mg/O \,7.65eV&1.75(.42)\\
Fe/O \,\, 7.87eV&1.63(.20)\\
Si/O \,\,\, 8.15eV&1.11(.30)\\
C/O \,\, 11.26eV&0.83(.20)\\
O/O \,\, 13.62eV&1.00(.11)\\
N/O \,\, 14.53eV&0.68(.27)\\
Ne/O \, 21.56eV&1.19(.17)\\
\hline
\end{tabular}
\end{center}
\end{table}


\begin{figure}
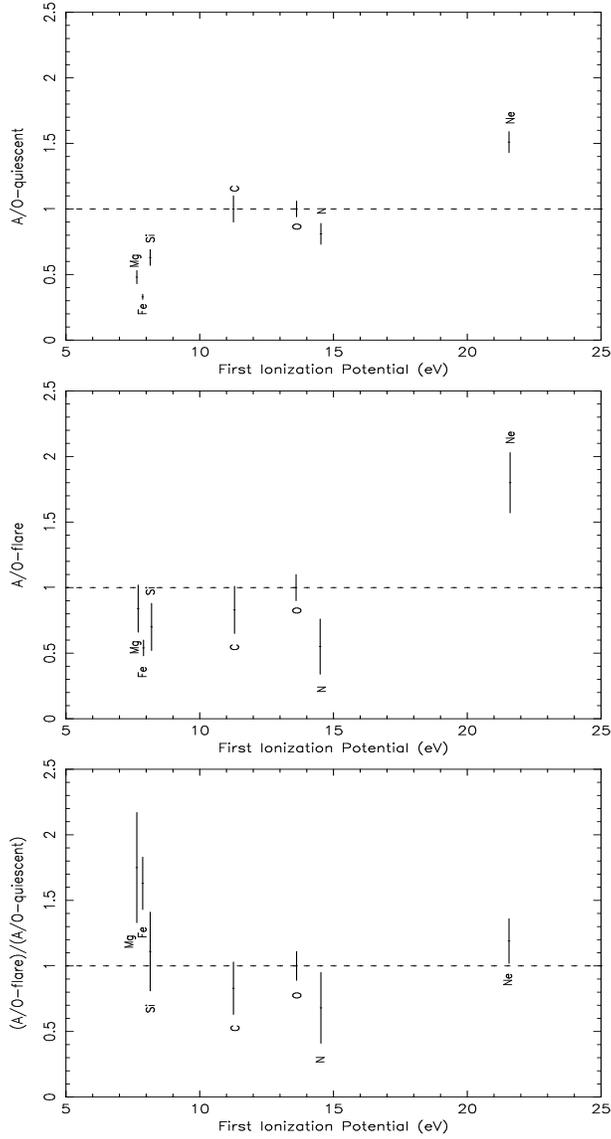

\vbox{
\center{\psfig{figure=h4290F4a.ps,angle=-90,height=5truecm,width=8.0truecm,clip=}}
\center{\psfig{figure=h4290F4b.ps,angle=-90,height=5truecm,width=8.0truecm,clip=}}
\center{\psfig{figure=h4290F4c.ps,angle=-90,height=5truecm,width=8.0truecm,clip=}}
\caption{Abundance ratio (A/O) versus First Ionization Potential, relative to solar photospheric values by
Anders \& Grevesse (1989) and Grevesse and Sauval (1998 and 1999) and with oxygen values from Allende Prieto \& Lambert (2001) for the
quiescent state (top) and the flaring state (middle). The bottom panel shows the ratio between the A/O of the flaring state and
the A/O of the quiescent state (Table~3).}
}
\end{figure}

 \begin{figure}[h]
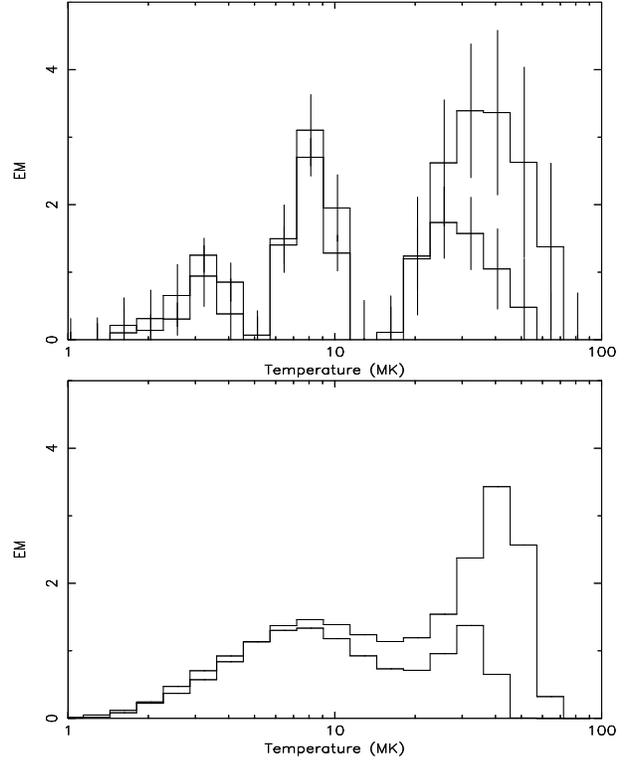

 \hbox
 {\psfig{figure=h4290F5a.ps,angle=-90,height=5truecm,width=8.0truecm,clip=}}
 {\psfig{figure=h4290F5b.ps,angle=-90,height=5truecm,width=8.0truecm,clip=}}
 \caption{DEM modeling of the flare spectrum (bold) and quiescent spectrum (weak) of AT Mic
in units $10^{51}$~cm$^{-3}$, using the regularization method (top panel). The bottom panel shows the same for the
polynomial method of order~9.}
\end{figure}


\subsection{DEM-modeling}
To show the smooth connection between the three separated temperature components of the multi-temperature fit (Table~2)
we have performed a DEM-modeling of the {\sc rgs} and {\sc epic-mos} 
spectra of AT Mic in the quiescent and flaring state, applying the regularization method and the polynomial method present 
in the {\sc spex}-code (see Kaastra et al. (1996b)). 
In this DEM-modeling the abundance ratios given in Table~2 have been used in combination with an O/H value of 1.57 for the
quiescent state and 1.39 for the flare state.
The emission measures $EM=n_{\rm e}n_{\rm H}V$ (per logarithmic temperature bin) 
of the corona of AT Mic are shown in Fig.~5ab for both states. The emission measure is distributed from $\sim$1 up to about 60~MK
for the flare state.
The DEM distribution agrees with the three temperature components from the multi-temperature fit if 
we take into account the uncertainties corresponding to the standard deviations. 

Table~2 and Fig.~5 show the increase of the emission measure and a shift
to higher temperatures for the hot EM (40~MK) during the flare state (bold) compared to the
quiescent state (weak). The emission measure below 10~MK is the same for the quiescent and flaring state.

\subsection{Individual lines}
The more prominent lines of the {\sc rgs} spectrum have been measured individually. Delta functions were folded through
the instrumental response matrix in order to derive integrated line fluxes. No additional width to the delta functions was needed
to fit the line shapes. A ``constant" level was adjusted to take into account the real continuum or the ``pseudo-continuum", created
by the overlap of numerous weak lines. 

The line fluxes are compared between the quiescent and flaring state (Table~4). Especially the flux difference of the 
Fe~XXV line at 1.84~\AA\ between the two states is clear. This line has a far higher flux in the flare state than in the quiescent
state. The same is true for other highly ionized (hot) ions, such as Si~XIV, Si~XIII, Fe~XIX, Fe~XVIII, and Fe~XVII. However, for these
ions the effect is less significant.

\begin{table}[bht]
\caption{Observed line fluxes of the quiescent and flare spectra of AT Mic.}
\begin{center}
\begin{tabular}[h]{l@{\ }l@{\ }|l@{\ }l@{\ }|l@{\ }l}
 \hline 
 \multicolumn{2}{c|}{{\sc quiescent}}&\multicolumn{2}{|c|}{\sc flaring}&\multicolumn{2}{|c}{{\sc Line ID}$^a$}\\ 
 \hline    
      $\lambda$(\AA) & Flux$^b$ &$\lambda$(\AA) & Flux$^b$ & $\lambda$(\AA)& Ion\\
      \hline 
       1.86(.05)$^c$         & 0.06(.3)    &  1.84(.5)      &0.42(.18)  &1.85  &   Fe~{\sc xxv}\\
       6.2(.2)$^c$   &  0.22(.12)&  6.2(.2)       &0.67(.42)   &6.182 &   Si~{\sc xiv}\\
       6.7(.2)$^c$   &  0.75(.13)&  6.7(.2)       &1.4(.8)    &6.7   &  Si~{\sc xiii}\\   
      12.136(.005)   &  4.7(.5) & 12.142(.009)       &5.3(1.2)    & 12.134              &  Ne~{\sc x}\\
      13.451(.005)   &  3.4(.4) & 13.442(.018)       &3.5(1.4)     & 13.447              &  Ne~{\sc ix}\\     
      13.533(.014)   &  1.0(.3) &13.521(.032)     & 2.4(.14)  & 13.521              &  Fe~{\sc xix}\\
                     &           &               &           & 13.553               &  Ne~{\sc ix}\\
                     &           &               &           & 13.670               &  Fe~{\sc xix}\\
      13.711(.014)   &  1.9(.3)  & 13.710(.018)    &1.8(.8)   & 13.700              &  Ne~{\sc ix}\\
      15.011(.009)   &  2.16(.31)& 15.012(.017)    &4.1(.9)     & 15.014             &  Fe~{\sc xvii}\\
      16.008(.009)   &  1.73(.29)  & 16.017(.009)     &2.7(.8)     & 16.007          &  O~{\sc viii}\\
                     &            &               &           & 16.078               &  Fe~{\sc xviii}\\
      16.763(.014)   &  1.20(.35)  & 16.791(.023)  &1.9(1.0)   & 16.775              &  Fe~{\sc xvii}\\
      17.050(f)      &  1.47(.37)  & 17.050(f)     &1.6(.8)   & 17.055              &  Fe~{\sc xvii}\\
      17.101(f)      &  1.30(.37)  & 17.101(f)     &1.6(.9)   & 17.100              &  Fe~{\sc xvii}\\
      18.655(.018)   &  0.53(.20)   & 18.645(0.46)    &0.5(.5)     & 18.628          &  O~{\sc vii}\\
      18.972(.007)   & 10.7(.7)   & 18.962(.006)     &11.2(1.5)   & 18.969           &  O~{\sc vii}I\\
      21.587(.009)   &  3.29(.61)  & 21.582(.018)    &3.9(1.0)    & 21.602           &  O~{\sc vii}\\
      21.789(.014)   &  0.64(.48) & 21.800(.024)    &1.4(.7)     & 21.801            &  O~{\sc vii}\\
      21.846(.037)   &  0.72(.50) &  ---          &---        & 21.845               &  O~{\sc vi}\\
      22.085(.009)   &  1.49(.18)  & 22.086(.034)  & 1.8(.6)   & 22.101              &  O~{\sc vii}\\
      24.781(.009)   &  1.07(.26) & 24.779(.028)  &1.1(.6)    & 24.781               &      N~{\sc vii}\\
      33.718(.007)   &  3.14(.53)  & 33.736(.018)    &2.8(1.2)   & 33.736            &  C~{\sc vi}\\
      \hline \\
    
\end{tabular}
 \end{center}
\begin{flushleft}
{
\begin{description}
\item $^a$ Line identification from  Kelly (1987) 
\item $^b$ Measured flux in 10$^{-4}$ photons/cm$^2$/s
\item $^c$ Wavelengths and fluxes from EPIC-MOS
\end{description}
}
\end{flushleft}  
\end{table}

\subsection{Density measurement}

\begin{figure}
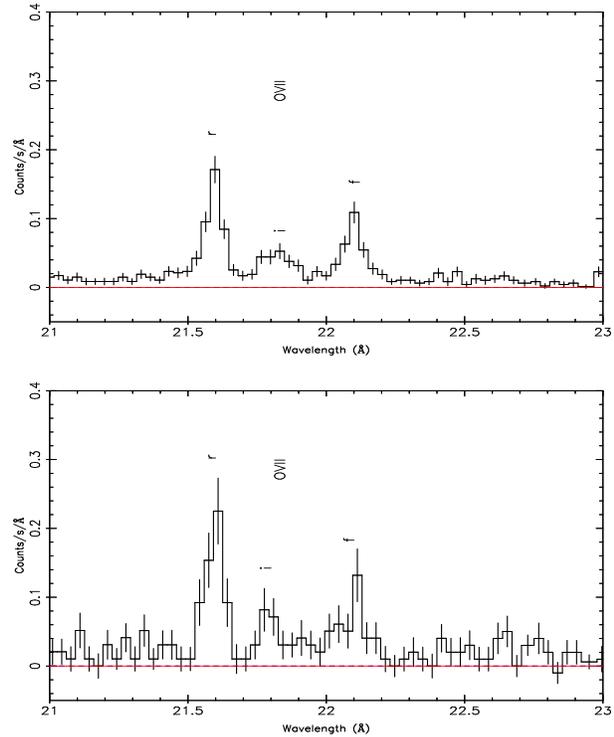

\hbox
{\psfig{figure=h4290F6a.ps,angle=-90,height=5truecm,width=8.0truecm,clip=}}
{\psfig{figure=h4290F6b.ps,angle=-90,height=5truecm,width=8.0truecm,clip=}}
\caption[]{ The O~{\sc vii} lines of AT Mic observed by {\sc rgs}1. The top panel shows the
quiescent state, while the lower panel shows the flaring state. }  
\end{figure} 

The three He-like transitions, consisting of the resonance line (r) 1s$^2$ $^1$S$_0$ - 1s2p $^1$P$_1$, the intercombination line (i) 
1s$^2$ $^1$S$_0$ - 1s2p $^3$P$_1$, and the forbidden line (f) 1s$^2$ $^1$S$_0$ - 1s2s $^3$S$_1$ are 
density- and temperature-dependent (e.g., Gabriel and Jordan 1969). At increasing electron density
the low-lying 1s2s $^3$S$_1$ level will be depopulated in favour of the 1s2p $^3$P$_1$ level, resulting in a decrease of the
intensity of the forbidden line and an increase of the intercombination line. This makes the ratio R = f/i a valuable diagnostic 
tool for densities. The He-like lines of oxygen are well separated, although the intercombination line in the quiescent
spectrum is affected by an innershell transition in O~{\sc vi}  (see Fig.~6). These 
lines at 21.602 \AA\ (r), 21.802 \AA\ (i) and 22.101 \AA\ (f) have been used to obtain temperature and density values
from the G- and R-ratio using the tables by Porquet et al. (2001). 
For representative temperatures around 3~MK, we derive electron densities of $3\pm 2 \times 10^{10}$~cm$^{-3}$ for the quiescent state
and $8^{+20}_{-7.5} \times 10^{10}$~cm$^{-3}$ for the flaring state. Within the errors these values agree with those given in Table~2.

\section{Conclusions}
The X-ray spectrum of AT Mic obtained by {\sc rgs} and {\sc epic-mos} is dominated by H- and He-like transitions 
of Ne, O, N, and C and by Fe~{\sc xvii} lines ({\sc rgs}) and by H- and He-like transitions of Si and Mg and Fe K-shell 
transitions ({\sc epic-mos}).

The spectra correspond to a quiescent and a flaring state as was concluded from the observed light curve.
The  dominant temperatures and emission measures cover a broad temperature range from 1 up to about 60 MK. The total 
emission measure in this temperature domain is $\sim 12.2\times10^{51}$~cm$^{-3}$ for the quiescent state and 
$\sim 19.5\times10^{51}$~cm$^{-3}$ for the flaring state. Both values are obtained from the ``fit weight data" fits.
  
From the {\sc O vii} lines a density of $\sim 2\times10^{10}$~cm$^{-3}$ is derived for the quiescent state and 
of $\sim 4\times10^{10}$~cm$^{-3}$ for the flaring state. The density of the flaring state is somewhat higher, 
but due to the poor statistics this is not significant.

Most line features of the {\sc rgs} spectra are produced in the temperature range from 3-10 MK, dominated by 
quiescent emission.  Lines from highly ionized atoms are enhanced in the flare state spectrum. 

The abundance pattern in AT Mic shows the depletion of low-FIP elements ($<$10~eV) relative to 
high-FIP elements in the quiescent state. This indicates the presence of a possible IFIP effect in this active star,
as has been shown for other active stars by Audard et al. (2002), Audard et al. (2003), Brinkman et al. (2001),
G\"udel et al. 2001, and G\"udel et al. (2002). 
However, in the flare state this effect appears to be suppressed. This may imply that during the flare material from
deeper layers with photosperic/chromospheric abundances is transported into the corona.

\begin{acknowledgements}

The SRON National Institute for Space Research is supported
financially by NWO.  
The PSI group acknowledges support from the Swiss National Science Foundation 
(grant 2100-049343).
MA acknowledges support from the Swiss National Science Foundation (fellowship 81EZ-67388).

\end{acknowledgements}

\end{document}